# A Tutorial on Event Detection using Social Media Data Analysis: Applications, Challenges, and Open Problems

Mohammadsepehr Karimiziarani
Department of Computer Science, University of Alabama, Tuscaloosa, AL, 35487, USA
Department of Civil, Environmental and Construction Engineering, University of Alabama, Tuscaloosa, AL, 35487, USA
Email: mkarimiziarani@crimson.ua.edu

**Abstract-In recent years, social media has become one of the most popular platforms for communication. These platforms allow users to report real-world incidents that might swiftly and widely circulate throughout the whole social network. A social event is a real-world incident that is documented on social media. Social gatherings could contain vital documentation of crisis scenarios. Monitoring and analyzing this rich content can produce information that is extraordinarily valuable and help people and organizations learn how to take action. In this paper, a survey on the potential benefits and applications of event detection with social media data analysis will be presented. Moreover, the critical challenges and the fundamental tradeoffs in event detection will be methodically investigated by monitoring social media stream. Then, fundamental open questions and possible research directions will be introduced.**

## I. Introduction

Online sharing platforms, social networks, and blogs have become increasingly popular as user-driven tools in online communication. Social media refers to all of these online sharing channels put together. [1]. Social media sites currently have more users than ever before. Social media is accessible practically anywhere and at any time, and users utilize it for a variety of things in addition to connecting with friends. A single post on a website can include insight about a person, an event, or anything happening in the world. Social media users' supplied data is used for a variety of things. First off, a lot of people use social media as a means of news since news organizations can reach more people on Twitter than they can through their websites alone. On the other side, users can exchange fresh knowledge using social media. Role of social media in event detection

By examining users' posts on social media sites, disasters can be predicted and identified using social media [2-4]. People occasionally spread news about occurrences before news organizations do. For instance, several people in the eastern US states reported that they learnt about the incident on Twitter even before they felt the earthquake in their area during the 2011 Virginia earthquake [5]. The data visualizations of the spread of the Virginia earthquake-related tweets exposed that the tweets traveled across the US faster than the actual earthquake [6, 7]. This makes Twitter and Facebook potential platforms for emergency notification. When traditional emergency warning systems failed, for instance, Twitter was used to communicate when a tsunami devastated Japan in 2011 [8]. In addition to its applications in event detection, the content shared online can give insight about people's opinion about various topics. Using this knowledge, researchers can gain information from users' posts. One use of such insights can appear in predicting how different people will vote in political elections.

### A. Motivation and objectives

Numerous analyses in data science have been conducted for extracting information from social media. This paper synthesizes the findings of many of those projects to discuss current applications, challenges, and open problems regarding the analysis of social media data. Our main goal in creating this article is to bring together a number of unique strategies that, when correctly coupled, should significantly improve event detection using social media data analysis. This research seeks to provide information about the current state of the field and open problems for areas of new research to researchers working in event detection with social media data analysis.

### B. *Real-time event detection*

In [9], the authors define an event as "A set of relations between social actors on a specific topic over a specified period". Events in social media analysis are denoted from apparent incidences of people, places, times and activities [10]. As introduced in [11], an event can be viewed as a single chapter of a big story, and social media data analysis can be beneficial for event detection. Events in social media analysis can only be understood from cumulative trend changes in the data stream [12]. Real-time event detection comprises the detection of remarkable happenings as they happen. The aforementioned events can be broad, e.g., when a flood occurs in Florida, or local events, e.g., traffic congestion in a particular street [13]. In this paper, we study the researches working on large scale events. As introduced in [14], large scale events can be inferred by three aspects: 1) These events are in large scale (numerous people experience the event), 2) They primarily affect people's daily life, and 3) Large-scale occurrences have both temporal and spatial characteristics (so that real-time location estimation is possible). These occasions include social gatherings such sizable parties, sporting occasions, art shows, mishaps, and political campaigns. Natural disasters including storms, torrential downpours, tornadoes, typhoons, hurricanes, cyclones, and earthquakes are also included.

### C. *Event detection technologies*

Both new event detection and retroactive event detection fall under the umbrella of event detection. Depending on the detection application, the technologies used for event detection in social media can be classified into three categories: 1) supervised event detection, 2) unsupervised event detection, and 3) hybrid event detection [15].

New event detection is an application of online clustering (unsupervised detection) which takes a high volume of text streams as input and then clusters the streams into topic clusters [16]. In this application of event detection, there is not a pre-determined event to be detected. For the most part, threshold-based clustering techniques have been used in new event detection. By defining a maximum similarity between fresh tweets and any of the current clusters, incremental clustering techniques, for example, are suitable for categorizing continuously created text. This approach provides an effective means of continuously streaming and clustering the events [17].

The majority of methods that concentrate on event detection use supervised learning strategies. Such techniques are more practical for detecting certain events because manually classifying many social media messages is a labor-intensive and time-consuming operation, and the supervised algorithms are utilized for that purpose.

While supervised classification and unsupervised clustering approaches are frequently used separately for event detection, novel event detection can make use of both techniques. Using a supervised classifier, the social media data is gathered, classified as relevant or irrelevant, and then clustering is used. This strategy is frequently called the hybrid approach [18]. Figure. 1 depicts a real-time novel event detection system's full workflow.

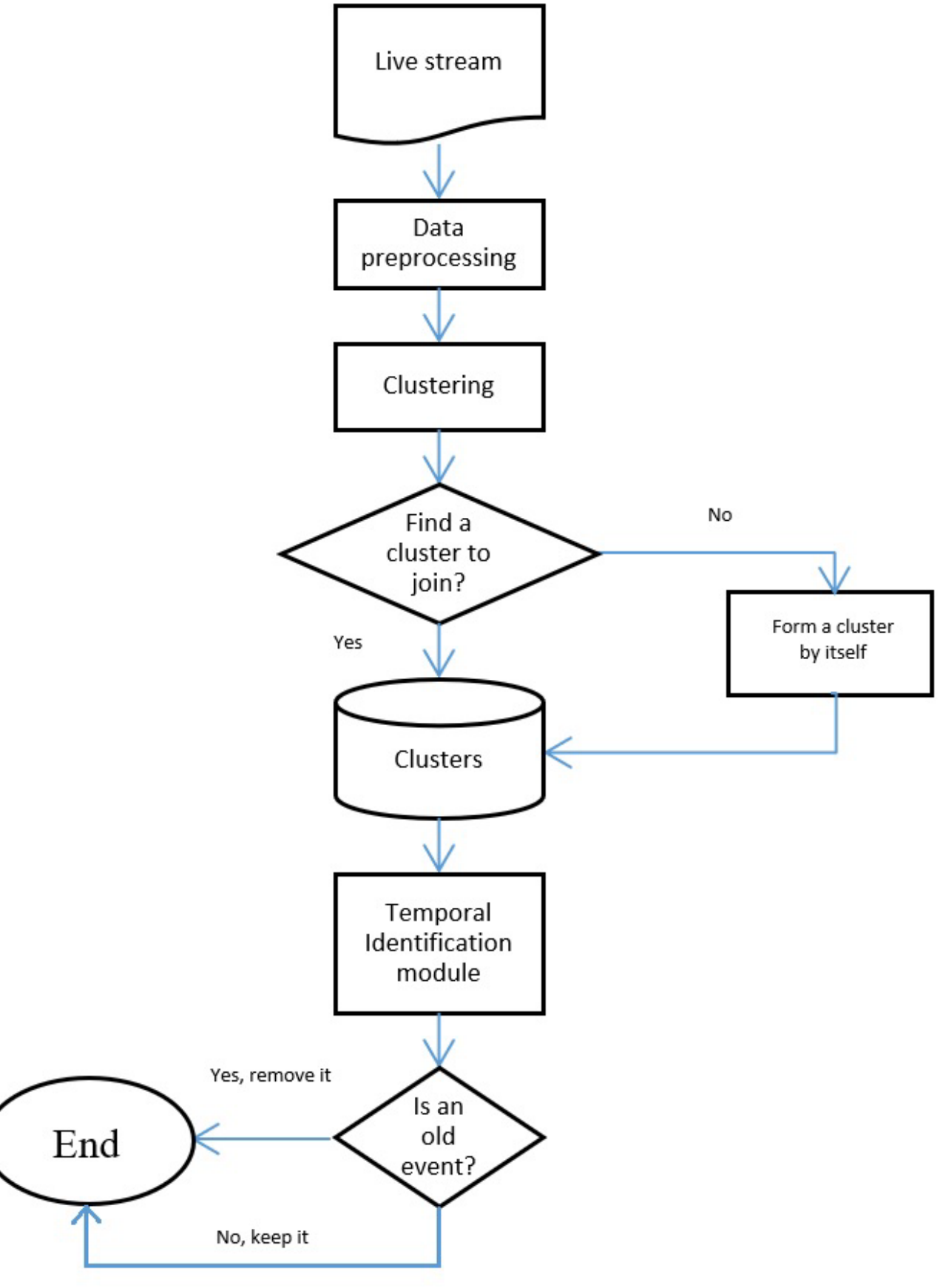

*Figure 1 Workflow of event detection from social media stream*

### D. *Limitations of prior work and our contribution*

There have been several reviews and surveys about event detection with social media analysis, few of which present a complete insight of the current applications along with open problems and challenges.

A few surveys emphasized on technologies and methodologies [19-22]. Social media data have great potential in applications, but only some of the favorite topics are summarized into surveys. [23] addressed attention on social media analytics for enterprises, and [24] examined the research and applications on mass emergency. Another set of researchers surveyed disaster response via social media analysis [25].

Although numerous surveys have been conducted, it would be much simpler for researchers to browse through current subjects and approaches if they had access to a tutorial that would provide them a better understanding of social media data analysis and applications. We made the decision to create a tutorial on social media analysis after examining the prior work.
To accomplish this, we address the following significant issues:

- A summary of possible event detection with social media applications is given in Section II. Future use cases and inspirational examples will be offered by these applications.
- The main research axes that will support the applications mentioned in Section II are outlined in Section III. To fully realize the potential of social media stream-based event detection, we explain for each study direction the research challenges and difficult open questions the solved.
- A list of the analytical frameworks that are anticipated to be crucial in the development of future social media stream-based event detection is presented in Section IV.
- In Section V, the article comes to a close with more information on this fascinating field of study.

## II. MOTIVATING APPLICATION USE CASES

To illustrate a clear picture of how social media data analysis can indeed be used for event detection, in this section, we overview several prospective applications for such a social media-centric event detection. The categories of natural disasters, traffic, politics, and outbreaks are by far the most prevalent in research, so the applications that are discussed here will focus on these four categories. In Figure 2, we provide an overview on the different types of events, and subfields. Most of the research that we reviewed focused on event detection in social media in English language. However, some papers did focus on the content from other languages.

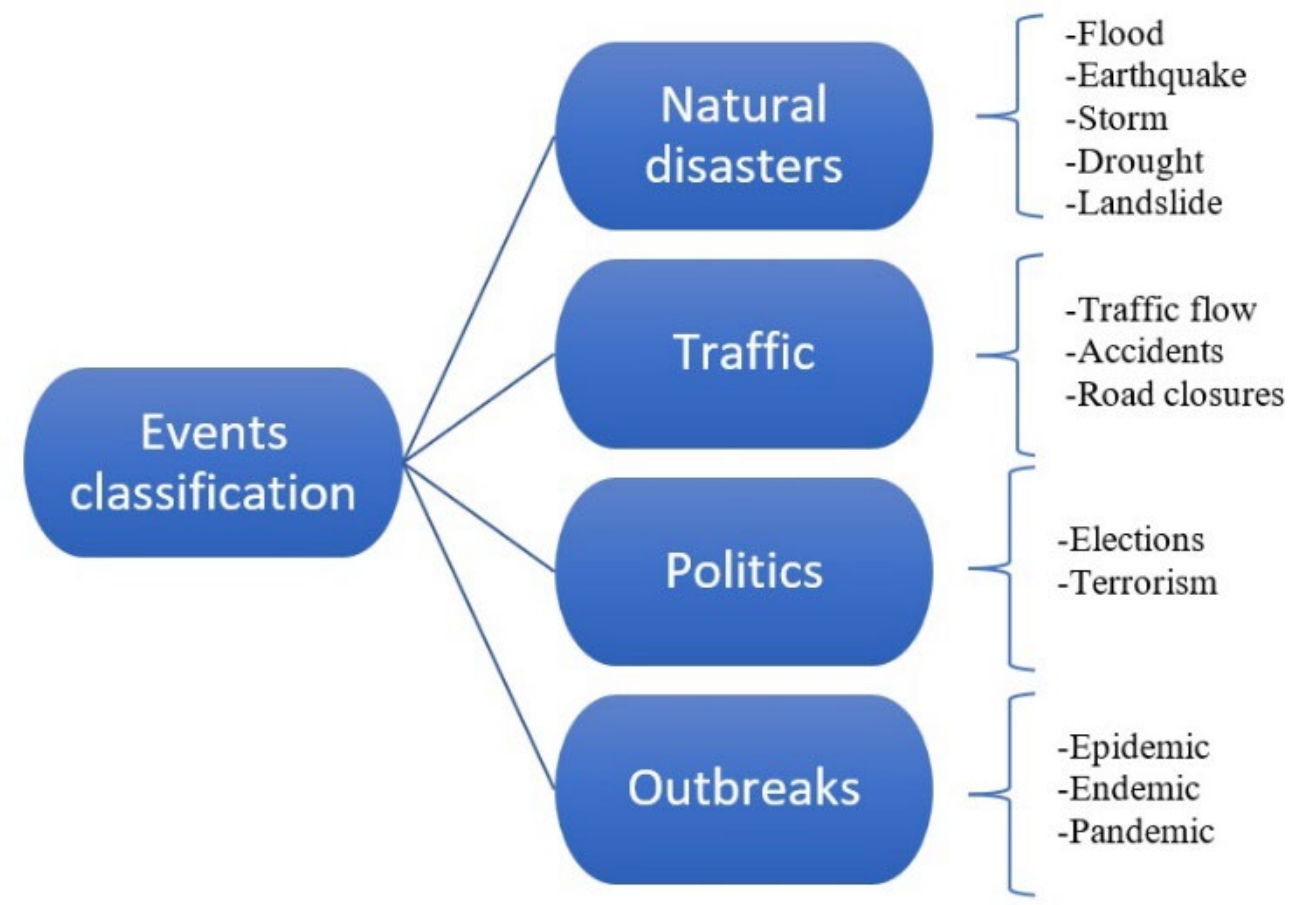

*Figure 2 Event classifications*

### A. *Natural disasters*

A disaster is "a potentially traumatic event that is collectively experienced, has an immediate beginning, and is time-limited," [26]. Natural catastrophes include phenomena like earthquakes, tsunamis, volcanic eruptions, floods, and other geologic occurrences that originate from the planet's natural processes. In many nations, natural catastrophes frequently have severe effects; they typically result in physical and economic losses as well as social, psychological, sociodemographic, economical, and political ones [27].

Typically, disasters are classified in three phases and can be studied through pre-event, event, and post-event phases [28].

Social networking has penetrated numerous industries, including the disaster relief sector [29]. Researchers, responders, and disaster planners are hopeful about the potential of social media to speed up improved disaster operations and communication [30]. Social media use amid an unplanned crisis could be incredibly beneficial. It could be done by posting information about the disaster and reading other users' or authorities' posts regarding the disaster and get vital information. Since smart electronic devices such as smart-phones and tablets may be used by citizens who can record and spread the information about the event, social media can play a bolder role during the event, even in the absence of professional news agencies [31].

In [32], the authors studied how people used Twitter in response to the 2017 Hurricane Harvey. Public interaction with social media platforms during the 2015 earthquake in Nepal was also analyzed by [33]. By examining these user-generated communications, it is possible to divide the content of social media during a natural disaster into five groups:

1. *Situational warnings and announcements*
   Situational details about the event are included in this category, such as floodwater levels and local traffic and road conditions. This category also includes announcements to seek shelter and emergency alerts such area evacuations.
2. *Support information*
   This group includes encouraging words like "free parking," "free emergency supplies," and "home repair consulting services."
3. *Help inquiries*
   This category includes messages asking for any kind of assistance, including food, water, medical supplies, volunteers, or transportation.
4. *Information inquiries*
   This category comprises general enquiries about disaster and disaster relief, such as phone numbers for relevant authorities, inquiries about the state of affairs in particular places, and inquiries about damage reimbursement.
5. *Other*
   This category includes all other postings, such as unspecific remarks, criticisms, and declarations of opinion.

Social media data analysis may assist the related authorities by delivering up-to-date information on the disaster such as the real-time needs of the civilians, damages caused by the disaster, road conditions, and much more valuable information. Government agencies and authorities can use this data to improve disaster relief methods in both during and after the disaster.

*B. Traffic Event Detection*

People frequently comment and complain about traffic on social media. Traffic events, including crashes, traffic jams, and road closures, are great concerns for both travelers and transportation management agencies. However, the detection and monitoring of traffic events remains a complicated task. Traditionally, transportation monitoring facilities such as loop detectors and cameras were applied on a relatively small scale, while the installation and maintenance of large-scale detection facilities are exorbitant. Crowdsourcing is emerging as a possible alternative method for identifying events and gathering traffic data as Twitter, Facebook, and other social media become more widely used.

Particularly, the event detection application in transportation concentrates on several topics.

*1) Traffic Accident Detection*

Traffic accidents breaking down the traffic flow and causing property damage, injuries, and fatalities is a worldwide issue in transportation research. Social media posts containing location and time of traffic accidents are utilized as individual sensors to detect the events.

Traffic congestion and surges in traffic always happen when traffic incidents occur which means accidents can cause traffic anomalies. [34-36] proposed real-time traffic event monitoring methods using social media to detect the incidents that are behind anomalies. A system for retrieving, processing, and categorizing tweets from Twitter that are connected to accidents was presented in [34]. The researchers next used the methodology to look at the detected occurrences' temporal and spatial coverage in the cities of Pittsburgh and Philadelphia.

A thorough method for traffic warnings and alerts based on the analysis of tweets is presented in [35]. With the help of their suggested semi-supervision, incremental learning, and tweet preprocessor, the proposed system implements multiple-level traffic warnings utilizing their twitter-LDA engine. However, it's still difficult for them to distinguish tweets about traffic from drivers with dependable precision. Supervised learning methods are commonly used in accident detection. [36] put into practice a strategy for crawling, pre-processing, and categorizing tweets utilizing a support vector machine and natural language processing. On their test dataset, the algorithm's overall accuracy for determining the sentiment of a tweet was 88.27%. Traffic management organizations can gain a better knowledge of how users view the network by using this data. To identify traffic-related microblogs from the Chinese social media platform Sina Weibo, [37] presented deep learning techniques. Convolutional neural networks (CNN), long short-term memory (LSTM), and its combination, LSTM-CNN, are utilized to categorize brief sentences from Sina Weibo. One-hot vector representation of words is applied for word embedding. Deep learning was used by [38] to identify traffic accidents from social media data in Northern Virginia and New York City, two major cities. To capture the association rules included in the accident-related tweets, researchers employed paired tokens. For incident identification, they used Deep Belief Network (DBN) and LSTM.

Tweets from official transportation management accounts are more accurate and convincing than individual Twitter accounts. Thus, [39] treated tweets separately as personal posts and organizational posts. For the two types of tweets, different dictionaries were developed to perform relevancy classification.

Instead of single incident detection, [40] concentrated on inspecting accident black spots, or places where traffic accidents tend to be concentrated.

[41] examined Twitter's potential for application in accident black spot prediction and confirmation. The accident-related tweets were grouped using density-based

spatial clustering of applications with noise (DBSCAN), and the researchers used the conclusive accident black spot data from an official tweet to analyze their approach. [42] examined the relationship between Twitter post concentration and traffic spikes brought on by social activity. Application of the supervised approach for classification using logistic regression.

*2) Traffic Congestion Detection*

Conventional traffic congestion monitoring relies on a significant number of traffic sensors or probe vehicles. The high cost limits the coverage and accuracy of the application. By exploring congestion related tweets or other posts, traffic information can be collected via various social medias.

Compared with traffic flow detectors (loop detectors, probe vehicles, etc.), tweets as sensors are not precise enough for accurate congestion detection and evaluation. Therefore, research about this topic concentrates on anomaly detection and mining relationships between traffic jams and Twitter posts. The researchers in [43] gathered historical information about traffic posts from particular cities and created a sentiment classifier to continuously track commuters' feelings. The information was utilized to evaluate and forecast traffic trends in a specific area. In order to efficiently integrate rich twitter information for Citywide Traffic Congestion Estimation, [44] suggested a linked matrix tensor factorization model (CTCE). [45] suggested an approach based on the analysis of Twitter tweets to anticipate the extent of traffic congestion. A C4.5 decision tree model for prediction was built using several types of twitter data. The information comprised tweets from chosen road traffic Twitter accounts, tweets with terms relating to traffic, and geotagged tweets whose volume indicated dense populations in specific places. By extracting sample phrases from social media posts made by users at the time the traffic anomaly occurred, [46] attempted to describe some of the discovered anomalies. This could offer information and understanding about traffic anomalies to both drivers and transportation officials. [47] developed a novel method for estimating traffic congestion that takes into account the timing and amount of comments and complaints made on social media websites.

*3) Other Traffic related Topics*

Traffic accidents and congestion are two major applications of event detection in the transportation area. Increased interest in social media analysis also brings about new applications, from travel information retrieval to road damage detection. [48] developed a systematic approach to examine social media posts near transit lines that talk about a public gathering and then proposed an optimization and prediction with hybrid loss function model to predict the passenger flow. [49] collected geo-tagged tweets with travel mode choices information posted in Melbourne metropolitan area for 23 weeks and investigated the travel modes information in the tweets. Researchers gathered and identified meteorological data from Twitter in [50]. They also investigated if using data from social media may increase the predictability of estimates of motorway traffic speed. [51] created the T-MAPS traffic information system specifically for the Manhattan area. The application incorporated route information to route recommendation and used tweets to convey the sentiment of routes. There are obviously many uses for traffic event detection using social media.

*4) Challenges*

Though traffic event detection is a promising field, there are a number of challenges that arise when trying to detect traffic events using social media. These are listed below.

There aren't any supporting context analyses or traffic-specific dictionaries to search for tweets about traffic. It is necessary to create a transportation ontology that defines an extensive list of concepts associated with traffic by utilizing collaborative intelligence and including input from non-experts.

Continuous updates cannot be met by the offline classifier trained using a batch of manually labeled geosocial media data.

The small datasets used for manually classifying the tweets in order to train a classifier are imbalanced in terms of positive and negative examples.

When collecting location information from tweets, natural language processing approaches overlook some terms linked to traffic. There are other abbreviations relating to the traffic domain that are frequently used to extract location mentions, like "av" for "avenue" and "st" for "street." With conventional natural language processing methods, these expressions are likely to be disregarded or unable to be associated with the appropriate terms.

How to properly correlate the sometimes-dissimilar locations of events and postings is a recurrent issue in research that use geosocial media data. Insufficient ground truth resources are also available to assess the traffic occurrences that have been detected.

## *C. Politics*

The use of social media for event detection in politics has increased significantly in recent years. Many scholars are examining how information is shared on social media during different political events, in addition to doing research on how to explicitly track political news events. In this lesson, we divided the social media event detection category of politics into two major sections. Election-related terrorism and politics are these.

*1) Elections*

Social media is used by many people to express their political opinions, including politicians who utilize it to stay in touch with their supporters and voters. It would be a remarkable accomplishment if social media could be used to anticipate elections, but sadly this is not the case right now [52]. The 2016 US presidential contenders and the election outcome were incorrectly predicted by one promising model that had 95.8% accuracy using test data [53]. Social media data can still be utilized to research politics even though accurate election prediction is not yet attainable.

Currently, social media is mostly a reactive tool in the context of elections [54]. In other words, online posts are often made in response to offline media forms [54]. However, there have been times that "social platforms might have provided information ahead of traditional news sources" [55]. For these reasons, monitoring social media can be used both to detect breaking news events and gauge the feelings that the public has toward different candidates during an election [55]. Data analysis techniques can be quite useful in tracking political elections.

*2) Terrorism*

Today's research focuses on the use of social media during terrorist strikes. Understanding how information is shared on social media after terrorist incidents is crucial for improving communication and ensuring people's safety. The majority of studies have therefore concentrated on the use of social media to effectively communicate during a terrorist attack and the response of citizens in such situations, even though there is some study on the detection of terror activities on social media [56, 57].

Social media takes an important role in disaster communication today [58]. In the scope of terrorist attacks, it is necessary to use it effectively to allow the spread of information critical to the events that are unfolding. In different cases, it has been found that the leading sources of information during terrorist attacks are often local news affiliates and law enforcements [59]. The 2013 terrorist attack at the Westgate Mall in Nairobi, Kenya is one event used as a case study to see how information can be effectively disseminated in the midst of an attack. To summarize, "Around noon on September 21, an unknown number of armed terrorists entered the Westgate shopping mall in Nairobi, Kenya and took control. During the following days, the Kenyan security forces tried to retake the mall using ground and air forces [60]." The examination of this incident offers insightful analysis of social media crisis communication trends. Updates were challenging to follow at first because there were numerous social media accounts being used by the officials to communicate information. However, by September 23, the government organizations that would post updates on social media began using the same hashtag in all their posts, so people were able to follow events more easily. Throughout this attack, social media was used as a crowdsourcing mechanism, with officials asking the public to share their messages, as well as asking those with information to post updates on what was going on. One challenge in the open communication of social media is that civilians sometimes shared information on social media about the activity of officials outside the mall, which in turn gave the terrorists awareness on the activity of the military and the police who were trying to retake the mall [60]. This event gave insight into benefits that social media can have for communication in terrorist attacks, but also problems that can arise while using it.

Another difficulty in using social media to communicate in terrorist events is that misinformation is easily spread through social media streams. During the Westgate attack, photos of the "attackers" were spread on Twitter, though the pictures were really of Kenyan armed forces [60]. Additionally, certain social media posts made following the 2013 Boston Marathon bombing misidentified the suspects, leading to an initial presumption that the wrong individuals were the bombers [61]. The rumor detection part of this study will address these challenges.

A further area of study with regard to terror attacks and event detection is the reaction of the public to terrorist events. One group of researchers used data from the terrorist event in Woolwich, London in 2013 to model the reaction to the attack from those using social media, to help those in charge of community safety to understand how the flow of information changes in the midst of an event like this [62]. They built models to predict the flow of information after an attack, including the period that different posts of various length and sentiment would continue to be shared [62]. It is evident that individuals, both those directly affected and those not part of the affected area, use social media to collectively make sense of terrorist events [63]. As a result, social media analysis can be used to monitor the long-term impacts that a terrorist attack has had on the local populace and, in certain situations, the nation that was the target of the attack [64]. Understanding how the public reacts to terrorist attacks and the effects these events have on the general population are both aided by social media data analysis.

## *D. Outbreaks*

Social media can be used effectively in two capacities about outbreaks. These are monitoring and surveillance, and situational awareness and communication surveillance [65].

First, data from social media can be analyzed so that outbreaks can be detected earlier. During an outbreak, there is a strong degree of correlation between discussion on social media of a disease and real-world epidemic activity [66]. While traditional reporting systems are highly accurate, data gathering can be quite slow, as healthcare professionals must report suspicions to a central health agency, and laboratory confirmations must be made [67]. For this reason, "detection of outbreaks through social media tracking appears to provide a timeliness advantage in a variety of infectious disease outbreak settings" [68]. This detection is not without challenges, however. Some of these challenges include the volume of content on social media and the noisiness of the data collected [69], as well as the potential for false alarms in detection [68]. However, these challenges can be addressed in order to make detection of outbreaks feasible.

Situational awareness is vital during an outbreak, and health officials must communicate to the public effectively to keep as many people healthy as possible. Analysis shows that the use of social media in epidemic communications offers multiple benefits, including the opportunity to disseminate information more quickly online than through traditional media outlets, as well as the ability to reduce the workload of crisis officials, as they can offer information more proactively, so that less requests for information will be made by the public [70]. Clearly, social media can be utilized to aid in communication during an outbreak. It can also be utilized to provide treatment and general health information. For example, both social media and recommendations from healthcare providers can influence parents' decisions on whether to vaccinate their children [71], so the effective use of social media can aid in community health outreach.

Similar to their use in politics, social media can be used in healthcare for detecting outbreaks and offering communication channels during outbreak events.

## III. RESEARCH DIRECTIONS, CHALLENGES, OPEN PROBLEMS, AND ADVANCES

An overview of the main areas of study that should be followed for event detection with social media data analysis will be provided in this part, which was motivated by the applications mentioned above. The main obstacles and unresolved issues for each study direction are described, and then the state-of-the-art approaches are discussed.

### *A. Candidate Tweets Filtering*

Even while social media APIs allow for the retrieval of a large number of posts, only a small portion of the data will be pertinent to the study question [24].
As the starting point for the subsequent analytic procedure, the precise extraction of traffic-related messages is crucial [72]. The two basic methods for extracting traffic-related data from raw tweets are the accounts-based method and the keywords-based method.

#### *1) Keywords-based approach*

These prospective tweets should often include one or more accident-related keywords from the research area, like "accident" or "crash." However, there hasn't been any agreement on a lexicon of words relating to the subject. As a result, the obtained data are always processed by either manually labeling the preprocessed tweets or filtering them based on frequently occurring phrases [73].
When collecting pertinent information, hashtags that are followed by post subjects can also be an alternative. This category describes tweets that exclusively contain specific weather-related information and utilize a hashtag like #weather to report a weather incident [74].

#### *2) Accounts-based approach*

Accounts can be categorized into general users and authority accounts.
General users constitute the majority of accounts and data sources. Due to the dispersibility and uncertainty, the data from general users can be messy and inaccurate. But registration and profile information of general users can provide the location information [44].

Real-time traffic updates are published on social media by official government accounts, such as those of the transportation or disaster management departments [75]. These users' information is more precise and provides specific temporal and spatial details about the incidents [44] compared to information from general users.

### *B. Data Pre-processing*

There are a number of issues with text mining because social media messages, and tweets in particular, are only so long. Language ambiguity, uncertainty, and brevity are a few of them. To find real-world occurrences, tweets must be cleaned before feature extraction [72].

Prior to feature extraction and modeling, this stage's goal is to process raw data using Natural Language Processing (NLP) methods. Tokenizing the tweets and deleting stop words and special characters are part of this procedure [76].

Natural language processing fundamentals are the foundation of pre-processing tweets. These include spelling corrections, replacing slang, stemming, lemmatizing, and POS tagging. They also include the removal of non-English words and duplicate posts (such as retweets), stop words, links to other Twitter accounts, and mentions of them.

Tokenization seeks to split the brief tweet text into individual tokens, and all of the letters are simultaneously normalized to lowercase. Accent marks, duplicate and non-English postings, connections to other Twitter accounts, and non-English posts are also removed. Stop words, including punctuation and non-alphanumeric characters, must be filtered as part of the cleaning procedure. Additionally, as personal posts frequently contain slang words and misspellings, it is necessary to make repairs. Finally, stemming—the removal of a word's suffix or prefix—occurs frequently. It is the act of reducing each word to its stem or root form.

### C. *Features Extraction*

Data must be accurately expressed by typically numerical vectors for various classification and clustering techniques [24]. Choosing a subset of features from the original documents is the basic goal of feature selection (FS). By maintaining the words with the highest scores in accordance with a preset metric for the word's relevance, FS is carried out.

Another aspect that aids classification is the frequency of terms in a tweet, which in certain cases approximates the relevance of the words. One of the most well-liked term-weighting algorithms takes into account both term frequency and inverse document frequency (tf-idf). The N-gram model, an essential and fundamental statistical model of natural language processing, was also utilized to process the few words in a tweet and provide characteristics. Stems, a bag of words, a lemma, POS and chunk features, a pattern recognizer, and a bag of tags are additional features that can be used.

### D. *Sentiment analysis*

Sentiment classification determines the viewpoint or attitude toward a circumstance or event by determining whether the textual item in question communicates a favorable or unfavorable perspective [77]. The main enabler technology for social media sites is sentiment classification. Political science, the social sciences, market research, and many other domains use content classification based on sentiment [78]. Though useful, sentiment classification is a difficult technical task. Since it is hard to read every social media post, automatic sentiment identification of text content has emerged as a critical task to enable quick response [79]. The fundamental difficulty with this task is that the areas in which the sentiment is expressed tend to vary often. Therefore, sentiment classifiers that can quickly adapt to new domains with a minimum supervision are required [80].

In the following section, recent studies on sentiment analysis will be discussed:

#### 1) *Retrofitting to Semantic Lexicons*

A strategy to improve word vectors utilizing relational data from semantic is suggested in [81]. It requires a vocabulary $V = w_1 \ldots, w_n$ , and its regarding word embeddings matrix $\hat{Q} = \{\hat{q}_1, \ldots, \hat{q}_n\}$, where each $\hat{q}_i$ is a vector for a word $w_i$ and an ontology Ω, that the authors characterize as an undirected graph (V, E) with a vertex for each of the word types and $(w_i, w_j) \in E \subseteq V \times V$. Authors attempted to learn the matrix $Q = \{q_1, \ldots, q_n\}$, such that $\hat{q}_i$ is both $\hat{q}_i$ and $q_j \vee j$ for *(i, j)* ∈ *E*. Thus, the objective function that should be minimized is:

$$\bar{\psi}(Q) = \sum_{i=1}^{n} \left[ \alpha_i ||q_i - \hat{q}_i||^2 + \Sigma_{(i,j) \in E} \beta_{i,j} \left|\left|q_i - q_j\right|\right|^2 \right],$$

the relative intensities of connections are controlled by α and β. The Paraphrase Database-XL version was utilized by the authors [82]. A collection of paraphrases called PPDB-XL is used as a semantic lexicon to improve the original vectors. Words in language A are deemed paraphrases if they are consistently translated to the same word in language B, and this dataset contains 8 million lexical paraphrases that are taken from text collections containing two different languages [84]. The Stanford Sentiment Treebank was used as the authors' second test subject [83]. The average word embeddings for a text are then used to train an L2-regularized logistic regression classifier, and improvements are discovered after retrofitting [84].

#### 2) *Supervised learning*

The most popular method for sentiment detection combines pre-trained supervised classification with word embeddings [84]. In supervised learning, the word embedding technique serves as the feature extractor for classification. Recurrent neural networks (RNNs), which include a memory state capable of learning long-distance relationships, are the LONG SHORT-TERM MEMORY network (LSTM) [85] and the GATED RECURRENT UNITS (GRU) [86]. They have been shown to be useful tools for text classification in a variety of ways [87]. [88] in combination with recurrent neural networks, [83] Glove vectors can be utilized to train the Stanford Sentiment Treebank. As the dataset is annotated for sentiment at each node of a parse tree, they train and test the annotated phrases. Different RNNs were introduced in [87] and [89] that can employ the marked nodes more effectively and produce results that are more accurate than those of ordinary RNNs. These models, however, require annotated parse trees, which may not always be present for other datasets. Convolutional Neural Networks (CNN) have achieved great results for text classification [90–92].

Skip-gram vectors may be tested on seven datasets, including the Stanford Sentiment Treebank [93], and utilized as input to a number of convolutional neural networks [116]. The most high-performance network, as compared to other configurations, can be thought of as a single layer CNN that has the ability to update the fundamental skip-gram vectors during training. The aforementioned methodologies have generally produced results that are more than sufficient on the majority of the current datasets, but they have not been contrasted with modified or joint training methods.

### *E. Geocoding*

After identifying harmful occurrences, several research collected the content's location from social media and geocoded the information using a range of toolsets for additional analysis. In order to detect traffic and disaster events, the geographic position data is crucial. Even though posts carry extensive geocode, it could also be quite noisy or not clearly available. Tagged GPS coordinates, user profiles, and tweet texts (such as place names and street names) are the three main categories of location data [72].

Although rich, the location data conveyed by posts is relatively noisy. There are typically three different types of location data.

#### *1) Coordinates tagged posts*

Only a small percentage of tweets include latitude and longitude information, and these are typically sent from smartphones with geotagging capabilities. This fraction is never higher than 0.1 percent in our experiments. While not necessarily the sites of the incidents, these coordinates are related to the places where individuals posted the TI tweets.

#### *2) Location information from profiles*

Some tweets are published by accounts whose public profiles include information on the city, country, and occasionally more specific business names and street addresses. Unfortunately, this kind of location data rarely indicates where events will be taking place.

#### *3) Location information from message texts*

Text messages may mention the names of places and areas of interest. Streets, landmarks, and direction information from the content can be deduced for the majority of data without geo-tags in order to extract position information and, if possible, map data to the GIS [44].

### *F. Rumor Detection*

Due to the prevalence of misleading information on social media, rumor detection presents a problem for event detection in social media data analysis. Additionally, "rumors communicated through social media can spread widely and cause mistrust, instability, and uncertainty" [94]. An item of circulating information whose factual status has not yet been confirmed at the time of publication is referred to as a rumor [95]. Feature identification is a crucial step in developing a rumor detection algorithm. A study on the features to employ to identify rumors on social media was conducted by Kwon et al. [96]. They discovered that linguistic characteristics consistently predict rumors. Initially, posts that contain words of skepticism or negation can be used to identify rumors; these characteristics consistently characterize rumors with great performance. User features are a different collection of features that are helpful in spotting rumors in their early stages. These demonstrate that users with little followers are ones who frequently write first about rumors, and the posts contain expressions of suspicion and doubt. These results led the researchers to suggest a linguistic and user feature-based rumor identification system. Various other categorization algorithms have also been suggested. One of them employs postings with questions or posts with skeptic-sounding language to identify rumors early on [97]. Another team of researchers discovered that leveraging implicit ties between social media postings that appear to be unrelated, including hashtags and web links, can be paired with the most advanced algorithms for enhanced identification of rumors with fewer posts [94]. Although there have been advancements in the field of rumor detection, this issue is still unresolved for researchers.

## IV. ANALYTICAL FRAMEWORKS

The analytical frameworks required to plan, examine, and maximize the usage of event detection for social media data analysis will be covered in this section.

#### *1) Machine learning*

By automatically learning from their surroundings and past experiences, machine learning (ML) enables systems to improve their performance. Computers can now identify patterns in data and group them into clusters thanks to machine learning [98-99]. Social media posts don't adhere to any standards, hence this approach is ideal for unstructured data. Typically, it combines text with images, sounds, video, and audio. The findings of such a study can provide important details on the writers of social media posts.

#### *2) Natural Language Processing (NLP)*

Artificial intelligence's subfield of Natural Language Processing. NLP enables computer systems to interact utilizing all varieties of natural human language. The term "natural language" has been used since the intention is to communicate with a computer or smart device using human languages like Persian or English rather than coding languages like Python or C.

| Challenges and Open problems | Key References | Covered Subjects |
| --- | --- | --- |
| Candidate Tweets Filtering | [24], [72], [73], [74], [44], [75] | • Keywords-based approach<br>• Accounts-based approach |
| Data Pre-processing | [72], [76] | • Language ambiguity<br>• Uncertainty<br>• Abbreviation<br>• Tokenization<br>• Normalization |
| Features Extraction | [24] | • Numerical vectors<br>• tf-idf<br>• N-gram |
| Sentiment detection | [77-93] | • Retrofitting to Semantic Lexicons<br>• Supervised learning |
| Geocoding | [72], [44] | • Coordinates tagged posts<br>• Location information from profiles<br>• Location information from message texts |
| Rumor Detection | [94-97] | • Feature identification<br>• Early detection of rumors |

*Table 1 Challenges and open problems for social media centric event detection*

The following characteristics of social media data are accessible sources of information that the whole public can access social, real-time, spatially tagged, emotional, neologisms, and credibility/rumors. These unstructured texts come in a variety of formats, are written in common English, and are authored by various authors in a variety of languages and writing styles. In addition, authors come from a diverse range of backgrounds and are not trained writers.

Creating effective techniques and algorithms that can draw out pertinent facts from a big volume of data in several languages is a difficult scientific task. A new type of data requires an adaptation of established NLP techniques in information extraction, automatic classification and clustering, automatic summarization, and machine translation.

Additionally, natural language processing provides insightful hints about the location, gender, age, and preferences of social media post authors. An NLP API's data can be used to segment customers based on actual data rather than assumptions or statistics.

# V. CONCLUDING REMARKS

We have presented research on event detection using social media data analysis in this tutorial. We have looked into the primary applications of social media data analysis for event detection. We have investigated important obstacles, applications, and core open issues for each situation. Along with interesting example results, we have also highlighted the key contemporary problems in event identification using social media data analysis. The mathematical methods and tools required for meeting event detection with social media data analysis requirements as well as analyzing event detection with social media data analysis were described in the meanwhile. We concluded by summarizing the major difficulties, unresolved issues, and crucial references for the analysis, optimization, and design of social media-centric event-detection systems.